\documentclass[%
 reprint,
superscriptaddress,
 amsmath,amssymb,
prl,
]{revtex4-2}

\usepackage{graphicx}% Include figure files
\usepackage{dcolumn}% Align table columns on decimal point
\usepackage{bm}% bold math
\usepackage{chemist}
\begin{document}
% \quickwordcount{manuscript}

%TC:ignore

\title{High energy density plasma mediated by collisionless resonance absorption inside dielectrics}% Force line breaks with \\

\author{Kazem Ardaneh}
\email{kazem.arrdaneh@gmail.com}
\author{Remi Meyer}
\author{Mostafa Hassan}
\author{Remo Giust}
\affiliation{FEMTO-ST Institute, Univ. Bourgogne Franche-Comt\'e, CNRS,15B avenue des Montboucons,\\ 25030, Besan\c{c}on Cedex, France
}%
\author{Chen Xie}
\affiliation{
%Ultrafast Laser Laboratory,
Key Laboratory of Opto-electronic Information Technology of Ministry of Education,\\
School of Precision Instruments and Opto-electronics Engineering,\\
Tianjin University, 300072 Tianjin, China\\}%Lines break automatically or can be forced with \\
\author{Benoit Morel}
\author{Ismail Ouadghiri-Idrissi}
\author{Luca Furfaro}
\author{Luc Froehly}
\affiliation{FEMTO-ST Institute, Univ. Bourgogne Franche-Comt\'e, CNRS,15B avenue des Montboucons,\\ 25030, Besan\c{c}on Cedex, France
}%
\author{Arnaud Couairon}
\affiliation{%
CPHT, CNRS, Ecole Polytechnique, Institut Polytechnique de Paris,\\ Route de Saclay, F-91128 Palaiseau,
France\\}%
\author{Guy Bonnaud} 
\affiliation{%
CEA, Centre de Paris-Saclay, DRF, Université Paris-Saclay,\\ 91191 Gif-sur-Yvette, France\\}%
\author{Francois Courvoisier}
\email{francois.courvoisier@femto-st.fr}
\affiliation{FEMTO-ST Institute, Univ. Bourgogne Franche-Comt\'e, CNRS,15B avenue des Montboucons,\\ 25030, Besan\c{c}on Cedex, France
}%

\date{\today}% It is always \today, today,
\begin{abstract}
We demonstrate for the first time to our knowledge the generation of overcritical plasma densities inside transparent solids over  long distances using femtosecond laser pulses. This opens new avenues for high energy density physics in confined geometry such as warm dense matter study or the synthesis of new material phases. We show both with experiments and first-principles simulations, that femtosecond conical interference via a Bessel beam creates a dense plasma rod with typically 100 nm diameter in sapphire. The interaction is in ideal conditions to trigger collisionless resonance absorption. This mechanism plays a primary role in the energy deposition process, yielding a plasma  with an energy density on the order of MJ/cm$^3$ and a length that can reach several cm using only tabletop femtosecond lasers.
\end{abstract}
%\begin{abstract}
%Generating large volumes of dense plasmas with femtosecond laser pulses is key to important applications such as the development of extreme UV sources, and for fundamental research such as the creation of warm dense matter, a high energy density state encountered in the core of astrophysical objects such as planetary cores and brown dwarfs.  A new route to generate such warm dense matter is within the bulk of transparent solids, because it cannot expand into vacuum, and because its relaxation can lead to the formation of new material phases. However, dense plasma generation in transparent solids is conventionally restricted to extremely small spherical volumes. Here we show both with experiments and first-principles simulations, that femtosecond conical interference via a Bessel beam creates a dense plasma rod with typically 100 nm diameter in sapphire under ideal conditions to trigger resonance absorption which is identified for the first time in the bulk of solid dielectrics. We show this mechanism plays a primary role in the energy deposition process, yielding an energy density of several MJ/kg over distances that can reach several cm using only tabletop femtosecond lasers.
%\end{abstract}

\maketitle
                            
%TC:endignore

\begin{figure*}[!ht]
    \centering
    \includegraphics[width=0.9\textwidth]{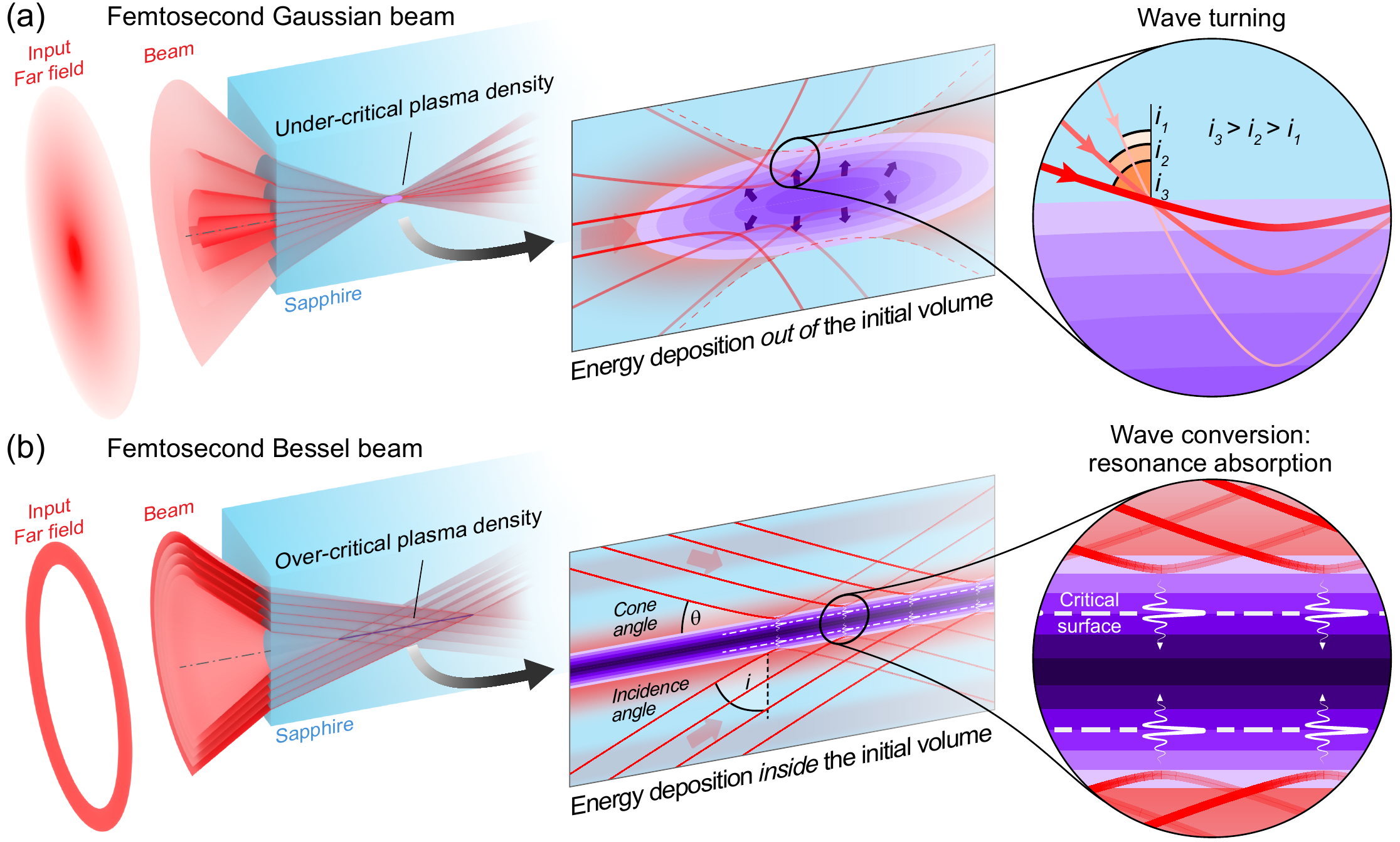}
    \caption{Laser-plasma interaction within dielectrics. (a) Gaussian beams are conventionally limited to create sub-critical plasmas in dielectrics because of the wave-turning phenomenon. Most of the waves composing the Gaussian beam impinge on the plasma with a very large incidence angle: these are deviated away well before they reach the plasma center. The plasma buildup expands the plasma {\it volume} but the density remains sub-critical. The radius of the turning point $r_{\rm t}$ is defined for an incidence angle $i$ by the density reached $n(r_{\rm t})=n_{\rm c}\cos^2i$, where $n_{\rm c}$ is the plasma critical density. (b) In a Bessel beam, the far-field k-space spectrum is narrow, the whole pulse energy impinges on the plasma with a relatively low incidence angle. The steep plasma density profile generated by the pulse onset makes resonance absorption possible, which transfers energy from the laser beam into the plasma. The energy transfer and ionization are therefore extremely localized on a sub-wavelength rod. The geometry of the interaction is longitudinally invariant.}
    %Laser-plasma interaction within dielectrics. (a) Gaussian beams are conventionally limited to create sub-critical plasmas in dielectrics because of the wave-turning phenomenon. Most of the waves composing the Gaussian beam impinge on the plasma with a very large incidence angle: these are deviated away well before they reach the plasma center. The plasma buildup expands the plasma {\it volume} but the density remains sub-critical. The radius of the turning point $r_{\rm t}$ is defined for an incidence angle $i$ by the density reached $n(r_{\rm t})=n_{\rm c}\cos^2i$, where $n_{\rm c}$ is the plasma critical density. (b) Because in a Bessel beam, the far-field k-space spectrum is narrow, the whole pulse energy impinges on the plasma with a relatively low incidence angle (importantly, for a stronger focusing the incidence angle onto the plasma surface is lower). The steep plasma density profile generated by the pulse onset makes resonance absorption possible, which transfers energy from the laser beam into the plasma. The energy transfer and ionization are therefore extremely localized on a sub-wavelength rod. The geometry of the interaction is longitudinally invariant so that over-critical plasmas can be generated over arbitrary distances -- limited only by the available input energy -- reaching several centimeters for mJ pulse energy level with the same cone angle as in here \cite{Meyer2019}.
    \label{figure1}
\end{figure*}

%\section{Introduction}
Extreme focusing of a femtosecond laser pulse with numerical apertures in excess of unity inside transparent materials generate a nanosphere of dense plasma, but the volume of the excited material is too small for X-ray investigation of warm dense matter \cite{Guillot1999,Saiz2008} or for the synthesis of new material phases \cite{Vailionis2011,Smillie2020}. Reducing the focusing strength enlarges the focal volume, but -- in contrast with laser interaction at the {\it surface} of solids -- the plasma density build-up {\it inside} the solid defocuses a Gaussian beam so that high plasma densities cannot be reached, even with higher pulse energy. The problem is illustrated in Fig.  \ref{figure1}({a}): after plasma generation on the optical axis by the initial part of a laser pulse, the trailing part of the pulse impinges on the plasma with a high angle of incidence $i$ with respect to the plasma surface normal. Instead of heating the plasma core, the optical pulse is deviated outwards to the regions with higher refractive index,  even for very low plasma densities. This is the wave-turning phenomenon \cite{kruer_1988,eliezer_2002} which presents a fundamental challenge. Importantly, the most intense components of a Gaussian beam, at the center of its far-field distribution, have largest incidence angles. They turn away at the largest distance from the plasma center [see circled inset in Fig.  \ref{figure1}({a})]. Due to the wave-turning, plasma generation continues outside  the initial volume and consequently the plasma remains sub-critical, {\it i.e.}, below the critical density at which the plasma permittivity turns to zero, and the deposited energy density remains low \cite{Bulgakova_2015,Naseri_2020}.

In this article, we report the creation of over-critical plasmas along quasi-arbitrary distances with femtosecond Bessel beams in sapphire, using only a moderately high numerical aperture (0.4). The plasma absorbs high amount of the pulse energy as a result of the resonance absorption mechanism, observed here for the first time in the bulk of solids. 

A zeroth-order Bessel beam is a circularly-symmetric interference field, so-called diffraction free \cite{Durnin87}, whose length is independent from the central lobe diameter [Fig.  \ref{figure1}({b})]. Its conical structure uniformly distributes energy by a transverse flow along the plasma created in the central core.  Resonance absorption has been widely studied within the context of inertial confinement fusion, by illuminating the surface of solids \cite{Gibbon_1992,Teubner_1993,Palastro_2018}. It occurs when a laser pulse obliquely impinges on a plasma ramp and the laser electric field has a component parallel to the plasma density gradient. In this condition, mode conversion resonantly drives electron plasma waves at the critical surface. This can  transfer up to 70\% of the laser energy into the plasma \cite{Denisov_1957,kruer_1988}. Mode conversion is efficient for the moderate incidence angle of Bessel beams on the plasma density ramp and generally inefficient in Gaussian beams because most of the energy is located on components at grazing incidence.

We determine the typical plasma profile by comparing complementary experimental diagnostics to first-principles Particle-In-Cell (PIC) simulations using the 3D massively parallel code EPOCH \cite{Arber_2015} because direct measurement of the plasma density inside the bulk cannot be performed via optical techniques when the plasma reaches the critical density at sub-wavelength scales. We demonstrate that in sapphire, over-critical densities are reached over a typical radius of 75-200 nm with typically 50\% absorption of the laser pulse energy. This is additionally confirmed by the observation of second harmonic from the bulk of a centrosymmetric dielectric for the first time to our knowledge. While conventional models of laser energy deposition hitherto have relied on collisional absorption, we show here that most of the absorption is due to collisionless resonance absorption. Our findings open a new route to develop high energy density physics in a confined geometry inside solids, but also open new perspectives for nonlinear optics with epsilon-near-zero materials \cite{Alam_2016}, synthesis of new material phases and laser material processing.

\begin{figure}[!htb]
	\centering
		\includegraphics[width=\columnwidth]{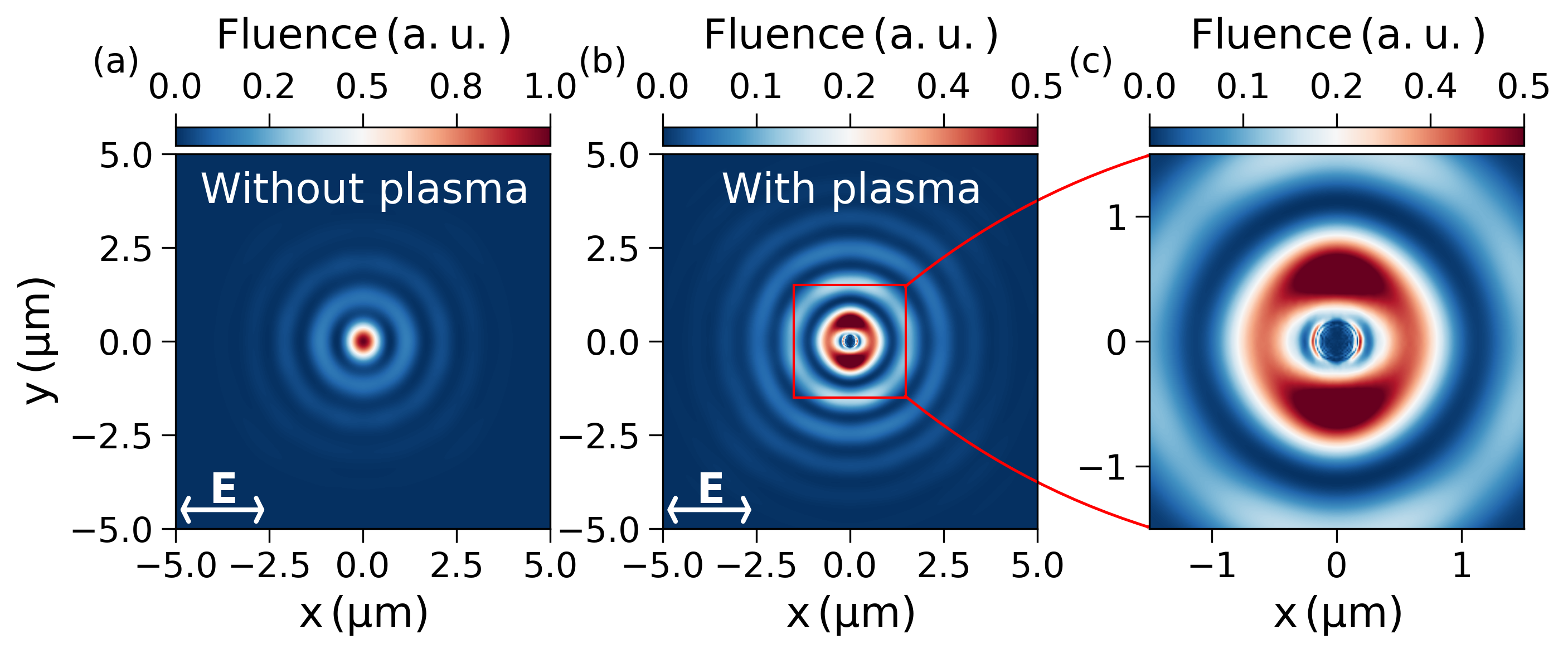}
	\caption{Near-field fluence distribution of the Bessel beam from PIC simulations. Fluence distribution in $xy-$plane at $z=7$~\textmu m (peak of the axial fluence profile), (a) in the absence of plasma and (b) in the presence of a circular plasma rod with a transverse Gaussian density profile. (c) Zoom-in view of the central part of (b). %(d), Same fluence distribution as in (c) but the fluence is calculated using the time integration of intensity $cB^2/2\mu_0$, instead of Poynting vector, to eliminate the contribution of electrostatic waves.
	}
	\label{Fig_cir}
\end{figure}
%%%%%%%%%%%%%%%%%%%%%%%%%%%% RESULTS
%\section{Bessel pulse interaction with a plasma}
The interaction of an ultrashort pulse with a dielectric medium  can be viewed as a two-stage process. In the first stage, field ionization and electron-impact ionization create a plasma, with which the trailing part of the laser pulse interacts during the second stage. Our experiments show that the femtosecond Bessel pulse generates a plasma over a length of 18~\textmu m with a typical diameter of a few hundreds of nanometers and absorbs approximately $50$\% of the 1~\textmu J pulse energy (similar plasma was produced over a length close to 1 cm with mJ pulse energy \cite{Meyer2019}). We point out that the ionization energy absorbed to produce an over-critical plasma of this size is $n_{\rm c} V U_{\rm g}$ where $n_{\rm c}$ is the critical density at 800~nm, $V\approx 1.5$~\textmu m$^3$ the plasma volume and $U_{\rm g}\approx10$~eV the bandgap of sapphire. The ionization energy is less than 0.04~\textmu J  which is negligible in comparison with the total absorbed energy. We therefore hypothesize that the first stage is very short compared to the pulse duration and we focus our modeling efforts on the second stage of the interaction. Our model does not take into account ionization in order to isolate the laser-plasma interaction.

We first qualitatively describe the physics of  Bessel beam interaction with plasma using numerical simulations. We have performed 3D PIC simulations to model the interaction of a 100~fs Bessel beam with a plasma rod. The pulse, with a central wavelength $\lambda=800$~nm, propagates along $z-$ direction and is polarized along the $x-$ direction \cite{Ardaneh_20}. In the absence of plasma, the intensity reaches $ 6\times 10^{14}\,{\rm W/cm^{2}}$, as in the experiments. The beam cone angle is $\theta=25^{\circ}$, which corresponds to an incidence angle $i= 65^{\circ}$ on the cylindrical surface of the plasma [Fig.  \ref{figure1}({b})]. In the central lobe the amplitude of the $z-$component of electric field is 25\% of the $x-$component, while the $y-$component remains negligible (2\%). We have presented in detail our numerical scheme in the Supplemental Material \cite{supplemental}.
%in Appendix \ref{Simulations}. 
In Fig.  \ref{Fig_cir}({a}), we show the fluence distribution, {\it i.e.}, the time-integration of the Poynting vector of the Bessel beam, in $xy-$plane at a propagation distance corresponding to the peak of axial fluence profile, without plasma. In Fig.  \ref{Fig_cir}({b}), the beam interacts with an over-critical circular plasma rod with a turning point radius of 274~nm. All lobes are shifted outwards by a distance of about 300~nm due to the wave-turning phenomenon (the slight discrepancy stems from the fact that the estimate of the turning point radius is made for a monochromatic plane wave). 

In the zoom-in view presented in Fig.  \ref{Fig_cir}({c}), two high-intensity structures appear inside the region of the initial central lobe. 
One is composed of two electromagnetic lobes outside the plasma ($y=\pm 0.5$~\textmu m), oriented perpendicular to the pulse polarization. They originate from the difference in permittivity between the plasma rod and the external medium \cite{Rajeev_2006,Liao_15}. This effect occurs even for sub-critical plasmas. It reduces the amplitude of the field in the polarization direction and amplifies it in the perpendicular direction.
The second structure is much more localized in the vicinity of the critical surface and shows two lobes along the polarization direction. It corresponds to resonantly driven electrostatic plasma waves at the critical surface. 
%These are mostly electrostatic. as they are absent in Fig.  \ref{Fig_cir}{d}. 
Therefore, this structure cannot be observed experimentally by beam imaging. This is why, in the following figures, we exclude the electrostatic field components from the simulated fluence distributions.

%\section{Near field experimental diagnostic: Upper limit of the turning point}

\begin{figure*}[tbp]%[tbp]
	\centering
		\includegraphics[width=\textwidth]{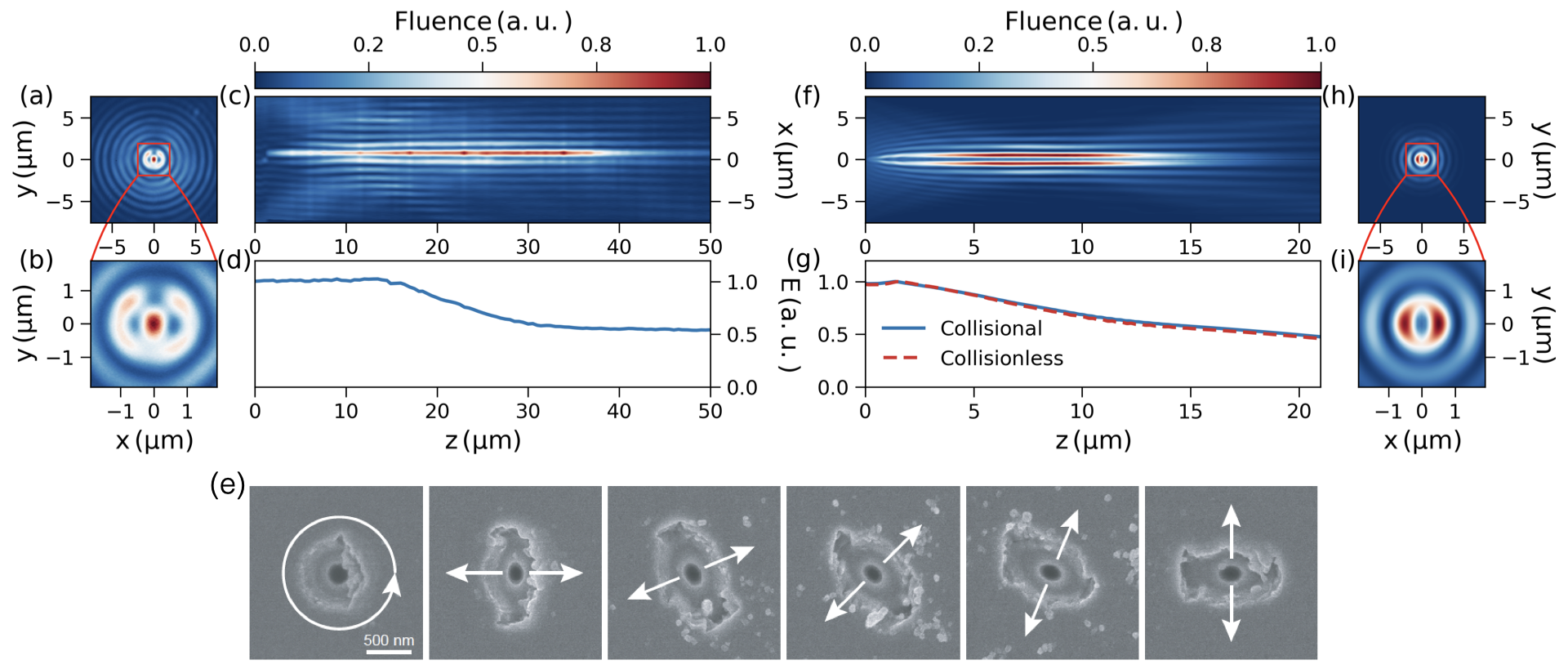}
	\caption{Near-field fluence distributions in sapphire: experiments (left) and PIC simulation (right). (a) and (h) Slices in $xy-$plane at $z=20$~\textmu m in (a) and $z=7$~\textmu m in (h). (b) and (i) Zoom-in views of (a), (h) respectively. (c) and (f) Slices of the fluence in the $zx-$plane ($y=0$). (d) and (g) Evolution of pulse energy as a function of propagation distance $z$ for collisional (solid blue line) and collisionless simulation (dashed red line).  (e)Scanning electron microscopy images of void channels produced in single shot illumination for different polarizations. The leftmost figure is for circular polarization and the channel cross-section is circular. In all other cases, the polarization is linear, oriented along the white arrow. %The orientation of the channel ellipticity follows the polarization direction.
	}
	\label{fig:NearField}
\end{figure*}

In Figs.  \ref{fig:NearField}({a})-\ref{fig:NearField}(d), we present experimental results of near field fluence imaging in 400 \textmu m thick C-cut sapphire \chemform{Al_2O_3}. They have been obtained by collecting the 100~fs laser pulse after its interaction with the sample, using a single shot beam sectioning technique \cite{Xie2015}. We have detailed our experimental techniques in the Supplemental Material \cite{supplemental}.
%in Appendix \ref{Setup}.
Figures \ref{fig:NearField}(a) and \ref{fig:NearField}(b) show the $xy-$cuts of the distribution at the peak of the axial profile ($z=20$~\textmu m).  Figure  \ref{fig:NearField}({c}) shows the fluence distribution in $zx-$plane while Fig.  \ref{fig:NearField}({d}) displays the evolution of the pulse energy as a function of propagation distance, measured with the imaging technique.

We observe a significant energy drop of nearly 50$\%$ over an 18~\textmu m segment of propagation, which corresponds to the region where plasma is generated. This is approximately the length of the void channel that is created in sapphire in single shot under similar laser illumination conditions \cite{Rapp_2016}. We independently checked that the energy drop is effectively equal to the total loss within the sample, confirming that energy deposition is localized in the high-intensity region, {\it i.e.}, not in an earlier stage of the propagation as it can be the case for Gaussian beams.

Importantly, we observe in Fig.  \ref{fig:NearField}({c}) that in the plasma region ($z= 14-32$~\textmu m), the  surrounding lobes negligibly shift with respect to those in the regions where plasma is absent ($z<14$~\textmu m, $z>32$ ~\textmu m). This sets the upper limit to the turning point radius $r_{\rm{t}}$ roughly to the experimental sensitivity associated with the width of the Bessel lobes, \textit{i.e.}, typically 500~nm. Given the incidence angle, the density at a distance $r_{\rm{t}}$ is less than $n_{\rm c}\cos^2i=0.086 n_{\rm{c}}$.

In Figs. \ref{fig:NearField}(a) and \ref{fig:NearField}(c), the shift due to wave turning is invisible within the central lobe. Because of the limited numerical aperture, the fine details of the fields inside the central lobe are not transferred to the imaging plane. To summarize, our experiments show that the onset of the Bessel pulse generates a plasma with a sub-wavelength diameter and that the trailing part of the laser pulse is significantly absorbed. 

% (An explanation could be that only a very low plasma density is generated, below $0.086\,n_{\rm{c}}$, but we show below that this cannot be the case because of high absorption).

We determine the typical plasma density profile  with a series of numerical simulations with different plasma profiles. We limited the turning radius to $r_{\rm t}=500\,{\rm nm}$ and chose the plasma transverse profile as a Gaussian distribution for which we adjust peak and width.
Collisional PIC simulations for sub-critical plasmas were showing an absorption of only a few percent, even for collision times as short as $0.1-2.5\,{\rm fs}$. As an example, with a peak density of $0.25~n_{\rm{c}}$, the turning radius has to exceed $900\,{\rm nm}$ to reach an absorption of 20\%. However, this would cause a remarkable radial shift of the Bessel beam lobes which is not found in our experiments. In contrast, only simulations with over-critical densities were  showing strong absorption in the range of $40-60$\% as in our experimental results.

 \begin{figure}[!htbp]
	\centering
		\includegraphics[width=\columnwidth]{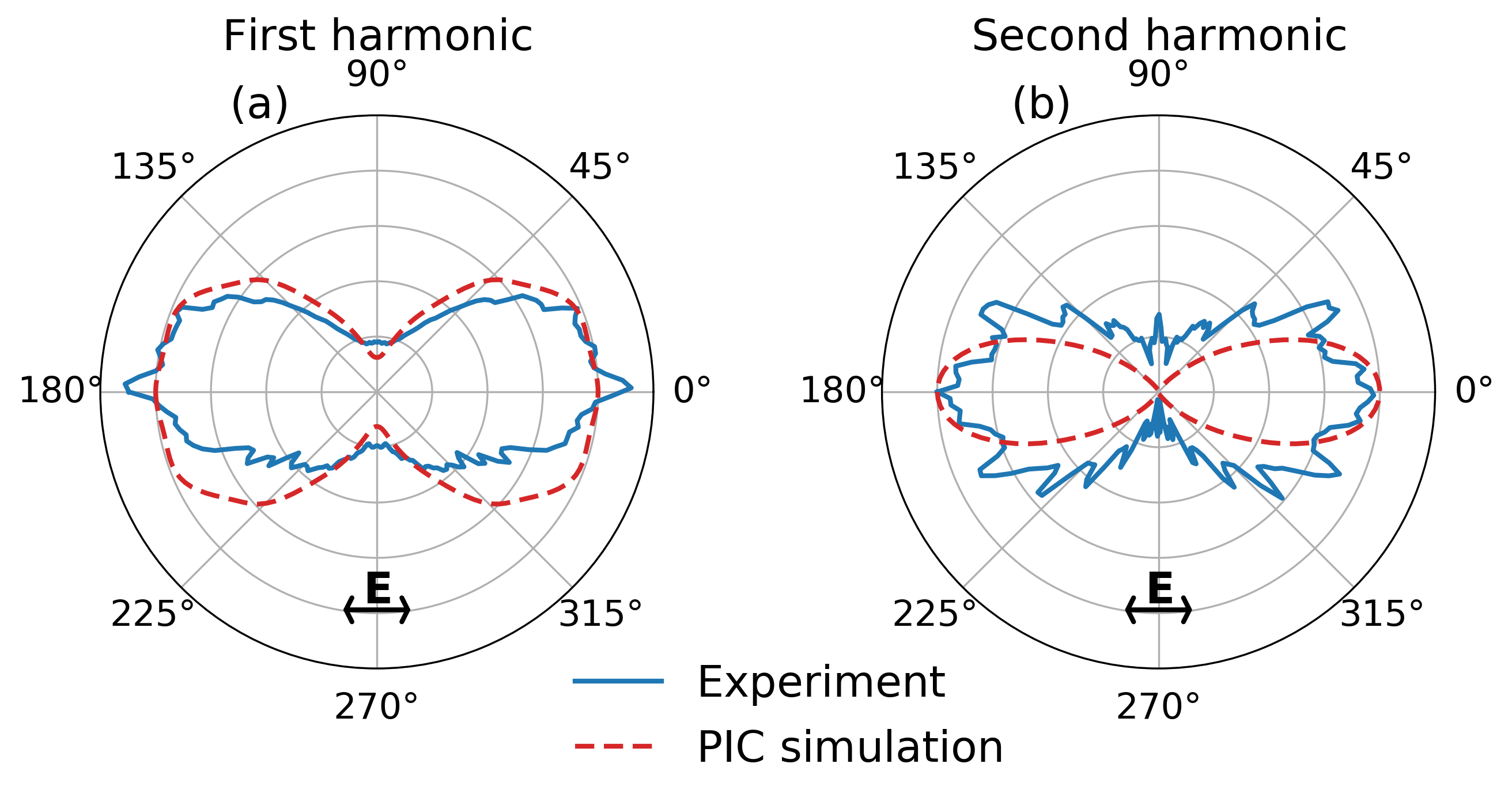}
	\caption{Angular distribution of far-field fluence for the first and second harmonics.The simulations share the same parameters as in Fig. \ref{fig:NearField}.}   	
	\label{fig:FarField}
\end{figure}

%{Far-field fluence distributions for the first and second harmonic.(a) Experimental far-field distribution of the laser pulse interaction with the sample. (b) Normalized angular distributions of the far-field distributions (experiment (a) in solid blue and simulation (c) in dashed red). (c) Numerical simulation, using the same parameters as in Fig. \ref{fig:NearField}. (d) Experimental far-field distribution of the emission in the spectral range $400 \pm 20$~nm, where the second harmonic appears as two intense lobes and black-body radiation as a uniform background filling the numerical aperture of the imaging. (e) Normalized angular distributions of the second harmonic in the experiment and simulation. (f) Simulated far-field distribution for the second harmonic.  In (a), (c), (d), (f), the axes are normalized by the radial component of the Bessel pulse, $k_{\rm r}^0=2\pi \sin\theta /\lambda$. The far-field simulations in (c), (f) show a broader pattern than in the experiments because of the limited numerical sampling.}
%\section{Elliptical plasma rod and quantitative comparisons}

The experimental near field profile  of Figs.  \ref{fig:NearField}(a) and \ref{fig:NearField}(b) is not cylindrically-symmetric.  With circular plasma rods, we were not able to reproduce the orientation of the asymmetries, neither the far-field distribution [see Fig.\ref{fig:FarField}(a)]. In contrast, we obtain a very good quantitative agreement with all our experimental results using elliptical plasma rods, having the major axis oriented perpendicular to the polarization (critical radius $r_{\rm c}=75-190$~nm along the polarization direction and $r_{\rm c}= 380$~nm perpendicular to the polarization). 

  The ellipticity can be explained by the field amplification in the first stage of field ionization, due to the difference in permittivity between the medium and the initially sub-critical plasma \cite{Rajeev_2006, Liao_15}. 
 We experimentally confirmed that the plasma is elliptical and elongated perpendicular to the polarization, by high-resolution imaging of voids formed in sapphire after the laser-induced micro-explosion, as shown for different input polarizations directions in Fig.  \ref{fig:NearField}({e}).

In Figs.  \ref{fig:NearField}({f})-\ref{fig:NearField}({i}), our simulations of the near field show negligible lobe shift of the Bessel beam and the asymmetries of the fluence distribution are oriented as in the experiment (we remind that the small dip observed in the central lobe of Figs.  \ref{fig:NearField}({h-i}) is filtered by the imaging operation). In the collisional plasma model [Fig.  \ref{fig:NearField}({g}) solid line], the pulse energy decreases linearly with a slope of $-0.03$~\textmu J/\textmu m and the total absorption is 54\%, both in good agreement with the experimental values. Overall, a steep plasma profile tends to increase the absorption factor and a narrow plasma density profile reduces the shift due to wave-turning. 
Importantly, the dashed curve in Fig.  \ref{fig:NearField}({g}) shows the result for a collisionless simulation. Since the difference with the collisional case is only a few percent (5 points), the main absorption mechanism is collisionless resonance absorption.

%\subsection{Far-field}

Our simulations show that the plasma diameter and its cross-section geometry strongly affect the far-field pattern.
%as shown in the Supplemental Material \cite{supplemental}. 
In the linear, absorption-free, regime, the far-field fluence of a Bessel beam is a homogeneous ring [Fig. \ref{figure1}(b)]. Supplementary Fig.  \ref{shg_supp}({a}) shows the experimental far-field fluence distribution collected after the interaction. Its angular distribution is illustrated in Fig.  \ref{fig:FarField}(a) (solid blue). It shows two lobes oriented in the direction of the input pulse polarization. 
The far-field from the simulation [Fig.  \ref{fig:FarField}({a}), red dashed line] is in quantitative agreement with our experimental data for the elliptical plasma rod. 

Overall, our simulations allowed us, up to here, to identify that the plasma is over-critical and with an elliptic cross-section. %Since the permittivity is zero at the critical density surface, nonlinear optical processes are possible {\it inside} the dielectric, on the plasma critical surface.

% \begin{figure*}[htbp]
%\begin{center}
%\includegraphics[width=\textwidth]{Figure5.png}
%\caption{Microphysics of the interaction between a femtosecond Bessel pulse and an elliptical sub-wavelength plasma rod. {(a)} Time evolution of the $x-$component of the electric field measured at $y=0$ and $z=7.5$~\textmu m in $xt-$space. {(b), (c), (d)} Electron phase-space at three different times $t-t_{\rm peak}=-78$, 28, $135\,{\rm{fs}}$. (e) electron energy distribution at the time $t-t_{\rm peak}=135\,{\rm{fs}}$. The over-plotted solid line in (a) shows the trajectory of a representative electron in which the color shows the kinetic energy of the electron. In (a), the dashed lines with a slope of $\approx 1\times10^7\,{\rm cm/s}$ show plasma expansion at the sound speed. The inset in (a) shows the electric field $E_x$ interpolated at the $(y_{\rm e},z_{\rm e})$, the coordinates of the electrons, and the over-plotted electron trajectory. The arrows in (c) indicate the critical surfaces located at $x=\pm 0.19$~\textmu m. The red dashed line (e) shows a Maxwellian distribution function at a temperature of $1.3\, {\rm eV}$ while the green dashed-dotted one shows one at a temperature of $70\, {\rm eV}$.}
%\label{circular_2}
%\end{center}
%\end{figure*}

%\section{Second harmonic generation}\label{Emission of Second Harmonic}

We experimentally observed second harmonic emission from the bulk of sapphire despite its zero second-order nonlinear susceptibility \cite{boyd_2008}. 
For this, we imaged the far-field fluence distribution in the spectral range 400$\pm$20~nm, as shown in Supplementary Fig.  \ref{shg_supp}({c}). Our experimental technique is described in the Supplemental Material \cite{supplemental}.
%Appendix \ref{Imaging of second harmonic generation}.
The distribution is composed of a low-intensity background originating from black-body radiation, and of a thin second harmonic emission at an angle nearly identical to the reflection of the pump. This reflection is superimposed to the transmission of the pump because of the conical interaction. 
%in the Appendix \ref{Discrete Fourier transform}
We show in Fig.  \ref{fig:FarField}(b) that our PIC simulations are in very good agreement with the experimental results.

In laser-plasma interaction, second-harmonic emission originates either from resonance absorption or from parametric decay instability \cite{erokhin1969}. In the latter case, the second harmonic polarization is perpendicular to the incident laser one \cite{Vinogradov_1973} while in the former case, it is parallel \cite{Auer_1979,Dragila_1982}, as in our experimental results.
Therefore, the second-harmonic diagnostic confirms that resonance absorption occurs.

In the simulations, the ratio of second harmonic to the fundamental intensity is $\sim 10^{-5}$. Using this ratio for Fig.  \ref{shg_supp}(c) and Stefan-Boltzmann law, one can estimate the plasma temperature on the order of 10~eV. This is consistent with our numerical framework and with the fact that Bessel beam-induced plasma relaxation forms warm dense matter in silicon dioxide \cite{Hoyo2020}.

We have demonstrated, for the first time to our knowledge, i) that  ultrafast Bessel beams can generate over-critical plasma within the bulk of sapphire, ii) that collisionless resonance absorption is a key process for energy deposition and iii)  second harmonic emission in single shot from the bulk of sapphire. Our findings therefore open a new route to generate high energy density matter inside solids. They answer a central question in the energy deposition by ultrafast pulses within the bulk of solids, which is particularly crucial in the field of laser micro-/nano- structuring of materials. Since the conical shape is propagation-invariant, our results can be easily extended to much longer propagation distances: a pulse energy of 1~mJ is enough to create a  1~cm long plasma in identical conditions as discussed above \cite{Meyer2019}. %Conversely, we expect that Bessel beams with smaller length or Bessel-like beams, such as the needle beam \cite{Wang_2008} as an example, can be efficiently used to outperform photoinscription in materials. The static field created during the pulse can find a number of applications to induce non-centrosymmetric structural changes in dielectrics for photonic components  \cite{Jiang_2018,Sugioka_2014}.

%The existence of an epsilon-near-zero material inside the bulk, due to the over-critical plasma, can find interesting applications for nonlinear photonics. It can particularly be used as a source of strong THz fields  \cite{Kampfrath_2013} parallel to the optical axis. Our simulations show a typical conversion efficiency of $10^{-3}$.  

 Nanoplasma generation with over-critical density inside solids opens interesting opportunities for nonlinear plasmonics  \cite{Kauranen_2012}, strong THz fields generation  \cite{Kampfrath_2013}  and extreme ultraviolet sources  \cite{Han_2016}. Controlling resonance absorption inside dielectrics provides useful new tools to drive solids into extreme thermodynamics pathways in a geometry where the material cannot expand into vacuum. The energy density deposited in our experiments is on the order of MJ/cm$^3$, which is a typical range for driving a transformation to warm dense matter  \cite{Denoeud_2014,Ernstorfer_2009}. Consequently, millijoule energies are enough to generate relatively large volumes of warm dense matter on the order of $1000$~\textmu $\rm m^3$. 
 
Our results therefore provide new avenues for generating relatively large volumes of new material phases and our configuration can be considered as a convenient new platform for the exploration of warm dense matter physics.
 
 %TC:ignore
\begin{acknowledgments}
K.A and R.M equally contributed, with the largest fraction of, respectively, simulation work and experimental work. Technical assistance by C. Billet and E. Dordor as well as fruitful discussions with J.M. Dudley and D. Brunner are gratefully acknowledged. 
We thank the EPOCH support team for help (https://cfsa-pmw.warwick.ac.uk), and French RENATECH network.

The authors acknowledge the financial supports of: European Research Council (ERC) 682032-PULSAR, Region Bourgogne-Franche-Comte and Agence Nationale de la Recherche (EQUIPEX+ SMARTLIGHT platform ANR-21-ESRE-0040), Labex ACTION ANR-11-LABX-0001-01, I-SITE BFC project (contract ANR-15-IDEX-0003), and the EIPHI Graduate School ANR-17-EURE-0002.
This work was granted access to the PRACE HPC resources at CINECA, Italy, under the Project "PULSARPIC" (PRA19\_4980), PRACE HPC resource at TGCC, France under the Project "PULSARPIC" (RA5614), HPC resource at TGCC, France under the projects A0070511001 and A0090511001, and M\'{e}socentre de Calcul de Franche-Comt\'{e}. 
\end{acknowledgments}
%TC:endignore

%\appendix

%\section{Absorption to reach the critical density}\label{Absorption to reach the critical density} 
%The ionization energy to create a plasma at critical density $n_{\rm c}$ inside sapphire is $n_{\rm c} V U_{\rm g}\approx$ 0.04~\textmu J where $n_{\rm c}$ is the critical density, $V$ the plasma volume (cylinder of length 18~\textmu m , $<500$~nm diameter) and $U_{\rm g}=10$~eV the band-gap of sapphire.

%\subsection{Bessel beam characteristics}\label{Bessel beam characteristics}
%The half-cone angle of the Bessel beam is $\theta=25^{\circ}$ in vacuum. 
%In the linear regime, the far-field of a Bessel beam is a ring distribution, peaked at $k^{0}_{\rm r}=2\pi\sin\theta/\lambda$, where $\lambda =800$~nm the central wavelength of the laser pulse.
%In the central lobe, for $\theta=25^{\circ}$,  the amplitude of the $z-$component of electric field is 25\% of the $x-$component, while the $y-$component remains negligible (2\%).

%\section{Turning point radius}\label{Turning point radius}
%For a given incidence angle $i$ on a plasma ramp, the turning point is defined by a density $n(r_{\rm t})=n_{\rm c}\cos^2i$, where $n_{\rm c}$ is the plasma critical density at which the permittivity turns to zero. At the turning point, the component of the wavevector parallel to the density gradient vanishes and reflection occurs. In our conditions, $n(r_{\rm t}) =0.086\,n_{c}$.

%\nocite{*}

\bibliography{manuscript2}% Produces the bibliography via BibTeX.
\clearpage
\newpage
\setcounter{section}{0}
\bfseries

\bfseries
\noindent {\large Supplemental material}
\vspace{1cm}

\mdseries
\renewcommand{\thefigure}{S.\arabic{figure}}
\renewcommand{\thetable}{S.\arabic{table}}
\renewcommand{\thepage}{S.\arabic{page}}
\setcounter{page}{1}
\setcounter{figure}{0}
\setcounter{table}{0}

%\newpage

\section{Particle-In-Cell simulations}\label{Simulations}

\subsection{Pulse injection}
Our numerical simulations model a plasma of electrons and ions in vacuum. We used ion-to-electron mass ratio $102\times 1836$ as the product of the sapphire molar mass and the proton mass.  We injected from the $z=0$ boundary a Gaussian pulse which propagates along the positive $z-$direction and is polarized along the $x-$direction. We applied a phase $\Phi(r)=-(2\pi/\lambda) r\sin\theta$ to the Gaussian beam, as it is the case experimentally using the spatial light modulator. The wavelength and cone angle are $\lambda=0.8$~\textmu m, $25 ^{\circ}$, respectively. The temporal profile of the beam is a Gaussian function with a width of $100\,{\rm fs}$ FWHM. The Gaussian beam waist is $w=10$~\textmu m. All simulations using a reduced cone angle, to simulate refraction, gave very similar results in terms of absorption, characteristics of near and far-field patterns. The Bessel beam length has been reduced to $\approx$ 18~\textmu m, by cropping the Gaussian beam window, to reduce the simulation run time. The peak intensity in the Bessel zone is $6\times 10^{14}\,{\rm W/cm^{2}}$ in absence of plasma, {\it i.e.}, on the same order of magnitude as in the experiments.

\subsection{Plasma parameters} \label{Plasma parameters} 
We have used circular and elliptical plasma rods. The axial density profile of the plasma rod follows $\tanh(z/z_0)$ with $z_0=1$~\textmu m. It shows a $z-$invariant density profile extending from $\approx$ 2~\textmu m away from the onset of the Bessel beam. The velocity distribution of both plasma species have been initialized with a Maxwellian distribution at $T_{\rm e}=T_{\rm i}$. We have run simulations for different initial plasma temperatures of $1\times10^{-3}$, 1 and 10~eV and we have obtained very similar results in terms of pulse absorption, structure of the fields, and hot electron population. All simulations results have included electron-ion collisions, although the collisionless case is almost indistinguishable.

\subsection{Resolution}\label{Resolution in PIC simulation}
We have used a computational box of $15\times15\times 30$~\textmu $\rm m^3$ volume and run the simulations up to $t_{\rm run}=320\,{\rm fs}$.  The spatial resolution of the grid for second-order FDTD is set so that $dxk^{0}_{\rm r}=dyk^{0}_{\rm r}=0.04$, and $dzk^{0}_{\rm z}=0.1$ where $k^{0}_{\rm r}=k_0\sin\theta$, $k^{0}_{\rm z}=k_0\cos\theta$. Using a fourth-order FDTD solver has not shown any significant change of the results. The temporal resolution is constrained by the Courant stability condition. We have used the convolutional perfectly matched layers boundary condition for the fields and open boundary condition for the particles. After testing the simulations with different numbers of particles per cell  from 32 to 128 to ensure energy conservation, we have chosen $32$ particles per cell per species (a total number of 550 million particles). We have used a first-order triangle shape function with 3 points to represent particles giving third-order particle weighting into the grid. We have obtained similar results using a third-order b-spline particle shape with 5 points. In the simulation of laser-plasma interaction using PIC simulation, the Debye length with typical value in the range $0.2-2$~ nm is not resolved by the grid cell, as it is the case in most simulations of laser-solid interaction  \cite{Arber_2015}. However, the energy is conserved using the energy conservative scheme for the integration of particle trajectory. In the EPOCH PIC code, the numerical heating is reported to grow as ${\rm d}T_{\rm eV}/{\rm d}t_{\rm ps}=\alpha_{\rm H}n_{23}^{3/2}\Delta x_{\rm nm}^2/n_{\rm ppc}$ where $T_{\rm eV}$ is the temperature in eV, $t_{\rm ps}$ time in ps, $n_{23}$ density in $10^{23}\,{\rm 1/cm^{3}}$, $\Delta x_{\rm nm}$ grid cell in nm, $n_{\rm ppc}$ number of particles per cell, and $\alpha_{\rm H}$ depends on the particle shape and current smoothing. For the triangular particle shape with current smoothing and the above resolutions $\alpha_{\rm H}=60$. Therefore, for the chosen resolution ${\rm d}T_{\rm eV}/{\rm d}t \sim 1\,{\rm eV/ps}$.

\subsection{Far-field distributions for first and second harmonics}\label{Discrete Fourier transform}
We have obtained the intensity spectrum $I(\omega,k_{\rm x},k_{\rm y})$ by performing a discrete Fourier transform on $B_{\rm x:y:z}(t,x,y)$ at a fixed propagation distance of $z=20$~\textmu m. The Fourier transformation has been performed on the recorded data for $\vert x \vert \leqslant 5$~\textmu m and $\vert y \vert \leqslant 5$~\textmu m. This is the reason for the width of the simulation far-fields in Figs.  (\ref{shg_supp}). The first and second harmonic radiation spectra are then obtained by filtering the power spectrum at the central frequency of the pulse $\omega_0$, and $2\omega_0$, respectively.

\section{Experimental setup}\label{Setup} 

\subsection{Fluence distributions}
We spatially shaped 100~fs laser pulses at a central wavelength of 800~nm. Using a spatial light modulator combined to a 2f-2f telecentric arrangement to produce a Bessel beam with cone angle 25$^{\circ}$ in air. The pulse duration of  100~fs has been characterized at the sample position. After the interaction, the pulse is collected with a $\times$100 microscope objective (Olympus MPLFLN) in a 2f-2f arrangement to image the pulse on a camera (Stingray F146B, 14 bits). The imaging system, with a numerical aperture of 0.9 is placed on an independent translation stage. The near-field fluence distributions of Fig.  (\ref{fig:NearField}) have been obtained by interrupting the propagation of the Bessel beam in the sample at the exit surface, as described in detail in reference  \cite{Xie2015}. The evolution of energy along the propagation direction has been measured by spatially integrating on the camera the far-field fluence at each propagation distance. The overall absorption in the sample has been confirmed by measuring the pulse energy on large-area photodiodes before and after the sample, as in reference  \cite{Hoyo2020}. 

%\subsection{Samples} \label{Samples} 
%The samples are 400 \textmu m thick C-cut sapphire \chemform{Al_2O_3}.

\subsection{Imaging of second harmonic generation}\label{Imaging of second harmonic generation}  
\begin{figure}[!ht]
\begin{center}
\includegraphics[width=0.7\columnwidth]{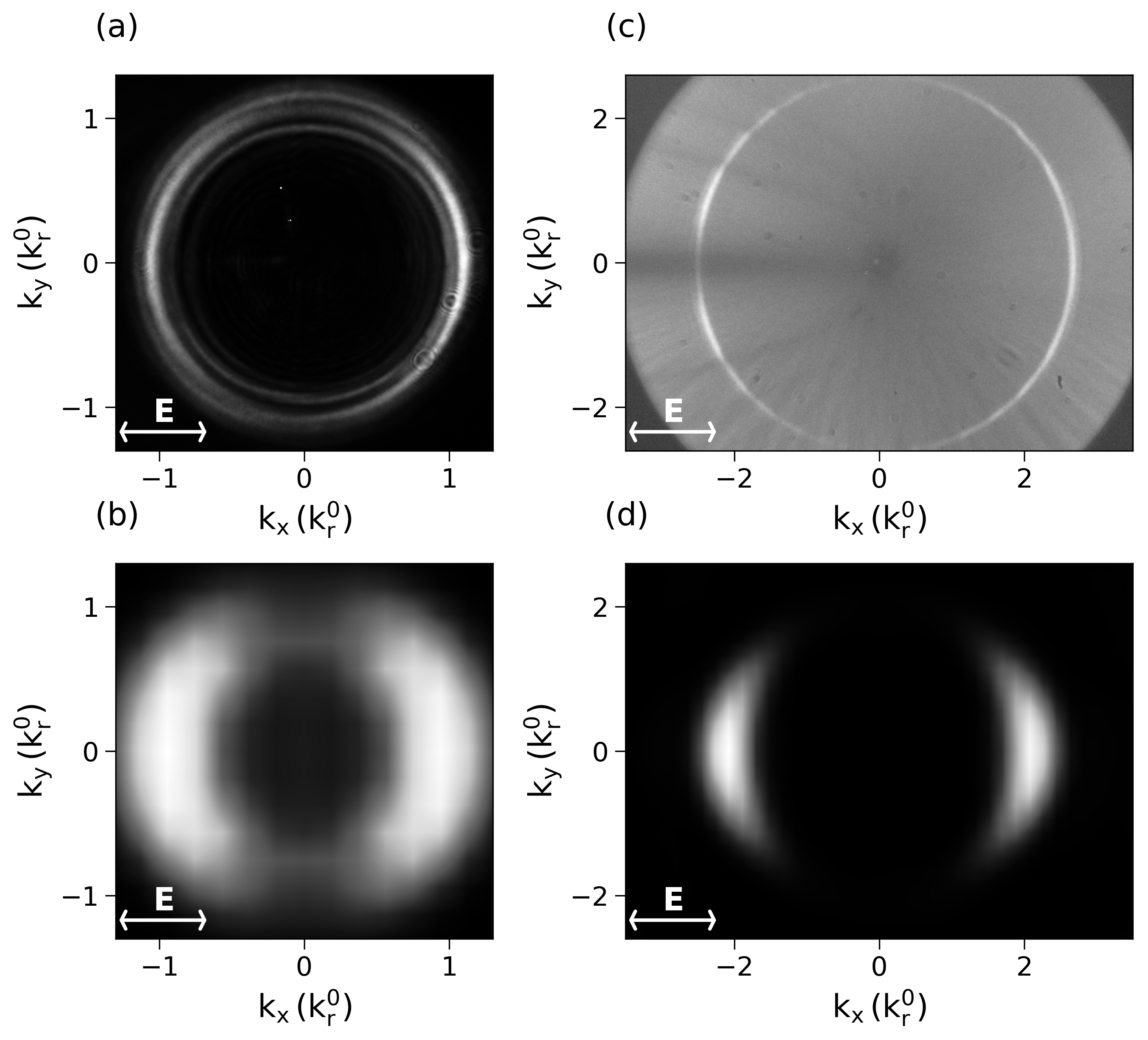}
\caption{Far-field fluence distributions for the first and second harmonic.
	Far-field distribution for first harmonic from (a) experiment  and (b) numerical simulation using the same parameters as in Fig. \ref{fig:NearField}. (c) Experimental far-field distribution of the emission in the spectral range $400 \pm 20$~nm and (d) simulated far-field distribution for the second harmonic.  All axes are normalized by the radial component of the Bessel pulse, $k_{\rm r}^0=2\pi \sin\theta /\lambda$.}
\label{shg_supp}
\end{center}
\end{figure}

Far-field images were recorded on a camera using an accumulation over 500 shots when the sample is continuously translated to separate the laser impacts by 5~\textmu m. The field was collected using a $\times$50 microscope objective (Olympus MPLFLN). The far-field distributions of Figs.  \ref{fig:FarField}(a) and \ref{fig:FarField}(b) have been symmetrized with respect to the $y-$axis. The measurement has been performed with the Bessel beam fully inside the bulk of sapphire. We remark that the second harmonic signal disappears when the input pulse is stretched to ps pulse duration. The signal is reduced when part of the beam crosses one of the air-dielectric interfaces, since the plasma length is reduced. The normalization of the axes of experimental measurements of Fig.  \ref{shg_supp} (a) and (c) has been performed using the pupil radius as the maximal numerical aperture (NA=0.8).

%\section*{Evolution of the far-field distribution for the principle and second harmonics for different plasma diameters}

%XXX here we need a small comment showing that if the plasma shrinks, the FF distribution is more circular. With large circular plasma, the lobes are perp to the polarization. With elliptical plasma, we get the correspondance to exp result.

%In our PIC simulations,  the laser interaction with plasmas of small critical radii ($\lesssim 72\,{\rm nm}$) produces a nearly circularly-symmetric far-field [Fig. (\ref{shg_fsh}) dashed red line]. In the case of wider circular plasma rods, the lobes were oriented perpendicular to the polarization direction [Fig. (\ref{shg_fsh}) solid blue]. The elliptical plasma rods oriented perpendicular to the laser polarization, however, can well reproduce the experimental far-fields [Fig. (\ref{shg_fsh}) dash-dotted green and dashed purple]. 

\end{document}